\documentclass[10pt,a4paper]{article}
\usepackage{amsmath}
\usepackage{amsfonts}
\usepackage{amssymb}
\usepackage[dvips]{graphicx}
\usepackage{bm}

\begin{document}

\title{\textbf{Responses of the Brans-Dicke field due to gravitational collapses}}
\author{Dong-il Hwang\footnote{enotsae@gmail.com}\;\; and Dong-han Yeom\footnote{innocent@muon.kaist.ac.kr} \\
\textit{Department of Physics, KAIST, Daejeon 305-701, Republic of Korea}}
\maketitle

\begin{abstract}
We study responses of the Brans-Dicke field due to gravitational collapses of scalar field pulses using numerical simulations.
Double-null formalism is employed to implement the numerical simulations.

If we supply a scalar field pulse, it will asymptotically form a black hole via dynamical interactions of the Brans-Dicke field.
Hence, we can observe the responses of the Brans-Dicke field by two different regions.
First, we observe the late time behaviors \textit{after} the gravitational collapse, which include formations of a singularity and an apparent horizon.
Second, we observe the fully dynamical behaviors \textit{during} the gravitational collapse and view the energy-momentum tensor components.

For the late time behaviors, if the Brans-Dicke coupling is greater (or smaller) than $-1.5$,
the Brans-Dicke field decreases (or increases) during the gravitational collapse.
Since the Brans-Dicke field should be relaxed to the asymptotic value with the elapse of time, the final apparent horizon becomes time-like (or space-like).

For the dynamical behaviors, we observed the energy-momentum tensors around $\omega\sim-1.5$.
If the Brans-Dicke coupling is greater than $-1.5$, the $T_{uu}$ component can be negative at the outside of the black hole.
This can allow an instantaneous inflating region during the gravitational collapse.
If the Brans-Dicke coupling is less than $-1.5$, the oscillation of the $T_{vv}$ component allows the apparent horizon to shrink.
This allows a combination that violates weak cosmic censorship.

Finally, we discuss the implications of the violation of the null energy condition and weak cosmic censorship.
\end{abstract}

\newpage

\tableofcontents

\newpage

\section{\label{sec:int}Introduction}

Einstein gravity is the most successful theory to describe gravitation.
However, if we include quantum effects, Einstein gravity cannot be a fundamental theory,
and hence a quantum theory of gravity is required.
This quantum gravity should be approximated by the Einstein gravity in the low energy limit,
but before approaching the Einstein gravity, there may be some effects that modify the Einstein gravity.
In fact, a theory of gravity does not definitively have to be Einstein gravity.
There can be many variations and we choose one theory via observations.
We then need to find possible modified gravity theories and check their possible implications.
If quantum gravity is approximated by a modified gravity theory, and if the theory has
experimental implications, it will be profoundly important to understand the nature of quantum gravity.

One of the most canonical modified gravity theory candidates is the Brans-Dicke theory \cite{Brans:1961sx}:
\begin{eqnarray}\label{eq:BD}
\mathcal{L} = \frac{1}{16\pi} \left( \phi R - \frac{\omega}{\phi}\phi_{;\mu}\phi_{;\nu}g^{\mu\nu} \right)
\end{eqnarray}
where $R$ is the Ricci scalar, $\phi$ is the Brans-Dicke field, and $\omega$ is the Brans-Dicke coupling constant.
Historically, the Brans-Dicke theory was suggested to implement Mach's principle.
The roll of the gravitation constant $G$ is replaced by the Brans-Dicke field $1/\phi$.
Since the Brans-Dicke field can be dynamic, the strength of gravity can be relatively different for each points.
It makes that the equivalence principle has more freedom than the original Einstein gravity,
since the equivalence principle holds for each points independently.
Then, intuitively, we can recover the Einstein gravity in the large $|\omega|$ limit,
since the field dynamics of the Brans-Dicke field will be highly suppressed in the limit.

Moreover, the Brans-Dicke gravity has other theoretical motivations.
Notably, it appears that it is the simplest form of modified gravity using a scalar field.
Therefore, some important intuitions of the Brans-Dicke theory can be applied to other possible modified gravity theories.
For example, we may apply similar intuitions to dilaton gravity, non-minimally coupled theory, etc.

The Brans-Dicke theory can be motivated by string theory inspired models.
For example, the Brans-Dicke theory is the weak coupling limit of dilaton gravity ($\omega=-1$), where the dilaton field
is a direct consequence of string theory \cite{Gasperini:2007zz}.
Also, the Brans-Dicke theory is a weak field limit of the Randall-Sundrum model \cite{Randall:1999ee},
where, on the positive tension brane, $\omega$ is sufficiently large; on the negative tension brane, $\omega \gtrsim -1.5$ \cite{Garriga:1999yh}\cite{Fujii:2003pa}.
Of course, the observed limitation of the coupling is $\omega > 4 \times 10^{4}$ \cite{Ber}; however, it is still meaningful to study other values of $\omega$ since they can be realized in the context of string theory,
and hence they can show properties which are allowed by string theory. For further discussions, see Appendix~A.

If the Brans-Dicke field is added to Einstein gravity, it can work as a source of an exotic matter.
In the context of cosmology, there have been some discussions that the Brans-Dicke field can be a candidate for dark matter or dark energy \cite{Setare:2006yj}.
In particular, it can be useful to study exotic matters that violates the null energy condition.
The theory can allow some strange geometry, e.g., wormholes \cite{Agnese:1995kd}.
Therefore, if we know how to deal with the Brans-Dicke field so as to obtain a sufficient amount of exotic matters,
then we can reasonably make an assumption of exotic matter,
since the exotic matter may thus be naturally obtained from a well-known and not artificial field.

Although the study of the Brans-Dicke field is important,
solutions to the Brans-Dicke theory are more difficult than those involving Einstein gravity.
Of course, we can find some analytic solutions.
However, to observe the real dynamics when a black hole is formed and evolved,
more detailed calculations are needed.
The collapsing matter will affect the background, and the background will affect the Brans-Dicke field.
The affected Brans-Dicke field will then give a back-reaction to the background geometry.
These processes are not easy to understand in an analytic way and it is necessary to use numerical calculations for their elucidation.

Previous papers have studied Brans-Dicke black holes using ADM formalism \cite{Scheel:1994yr}.
However, in this paper, we use double-null formalism \cite{doublenull}\cite{Yeom1}\cite{Yeom2} for the numerical calculations \cite{Chiba:1996zu}\cite{Avelino:2009vv},
thus making it easier to study the causal structures and also to plot all possible fields including the energy-momentum tensors.

The purpose of this paper is to study responses of the Brans-Dicke field during the gravitational collapse of a matter.
We prepare a flat background with a pulse of a normal scalar matter field and we then observe the response of the Brans-Dicke field.
We thereupon observe that dynamical responses of the Brans-Dicke field are highly non-trivial.

This paper is organized as follows:
in Section~\ref{sec:mod}, we discuss the details of our numerical simulations;
in Section~\ref{sec:res}, we discuss the results of our simulations and report some interesting observations;
in Section~\ref{sec:inter}, we document observations in terms of two issues (late time behaviors and dynamical behaviors) and
discuss their physical interpretations;
and in Section~\ref{sec:con}, we summarize the possible causal structures and discuss the physical meanings and implications.

\section{\label{sec:mod}Model for Brans-Dicke theory}

\subsection{\label{sec:The}Brans-Dicke theory}

The Lagrangian of the Brans-Dicke theory with a scalar field becomes \cite{Brans:1961sx}
\begin{eqnarray}\label{eq:BDscalar}
\mathcal{L} = \frac{1}{16\pi} \left( \phi R - \frac{\omega}{\phi}\phi_{;\mu}\phi_{;\nu}g^{\mu\nu} \right) - \frac{1}{2}\Phi_{;\mu}\Phi_{;\nu}g^{\mu\nu} -V(\Phi)
\end{eqnarray}
where $R$ is the Ricci scalar, $\phi$ is the Brans-Dicke field, and $\Phi$ is a minimally coupled scalar field with potential $V(\Phi)$.
Here, $\omega$ is the Brans-Dicke coupling constant which is a free parameter of the theory.

The Einstein equation becomes as follows:
\begin{eqnarray}\label{eq:Einstein}
G_{\mu\nu} = 8 \pi T^{\mathrm{BD}}_{\mu\nu} + 8 \pi \frac{T^{\Phi}_{\mu\nu}}{\phi} \equiv 8 \pi T_{\mu\nu},
\end{eqnarray}
where the Brans-Dicke part of the energy-momentum tensors are
\begin{eqnarray}\label{eq:T_BD}
T^{\mathrm{BD}}_{\mu\nu} = \frac{1}{8\pi \phi} \left(-g_{\mu\nu}\phi_{;\rho \sigma}g^{\rho\sigma}+\phi_{;\mu\nu}\right)
+ \frac{\omega}{8\pi \phi^{2}} \left(\phi_{;\mu}\phi_{;\nu}-\frac{1}{2}g_{\mu\nu}\phi_{;\rho}\phi_{;\sigma}g^{\rho\sigma}\right)
\end{eqnarray}
and the matter part of the energy-momentum tensors are
\begin{eqnarray}\label{eq:T_scalar}
T^{\mathrm{\Phi}}_{\mu\nu} = \Phi_{;\mu}\Phi_{;\nu}-\frac{1}{2}g_{\mu\nu}\Phi_{;\rho}\Phi_{;\sigma}g^{\rho\sigma} -g_{\mu\nu} V(\Phi).
\end{eqnarray}

The field equations are as follows:
\begin{eqnarray}
\label{eq:phi}\phi_{;\mu\nu}g^{\mu\nu}-\frac{8\pi}{3+2\omega}T^{\Phi} &=&0, \\
\label{eq:Phi}\Phi_{;\mu\nu}g^{\mu\nu}-V'(\Phi)&=&0,
\end{eqnarray}
where
\begin{eqnarray}\label{eq:T}
T^{\mathrm{\Phi}} = {T^{\mathrm{\Phi}}}^{\mu}_{\mu}.
\end{eqnarray}

\subsection{\label{sec:imp}Implementation in double-null formalism}

We use the double-null coordinates
\begin{eqnarray}\label{eq:doublenull}
ds^{2} = -\alpha^{2}(u,v) du dv + r^{2}(u,v) d\Omega^{2},
\end{eqnarray}
assuming spherical symmetry, where $u$ is the retarded time, $v$ is the advanced time, and $\theta$ and $\varphi$ are angular coordinates.

We follow the notation of \cite{Hamade:1995ce}\cite{doublenull}\cite{Yeom1}\cite{Yeom2}: the metric function $\alpha$, the radial function $r$, the Brans-Dicke field $\phi$, and a scalar field $S \equiv \sqrt{4\pi} \Phi$, and define
\begin{eqnarray}\label{eq:conventions}
h \equiv \frac{\alpha_{,u}}{\alpha},\quad d \equiv \frac{\alpha_{,v}}{\alpha},\quad f \equiv r_{,u},\quad g \equiv r_{,v},\quad w \equiv \phi_{,u},\quad z \equiv \phi_{,v}, \quad W \equiv S_{,u},\quad Z \equiv S_{,v}.
\end{eqnarray}

The Einstein tensors are then given as follows:
\begin{eqnarray}
\label{eq:Guu}G_{uu} &=& -\frac{2}{r} \left(f_{,u}-2fh \right),\\
\label{eq:Guv}G_{uv} &=& \frac{1}{2r^{2}} \left( 4 rf_{,v} + \alpha^{2} + 4fg \right),\\
\label{eq:Gvv}G_{vv} &=& -\frac{2}{r} \left(g_{,v}-2gd \right),\\
\label{eq:Gthth}G_{\theta\theta} &=& -4\frac{r^{2}}{\alpha^{2}} \left(d_{,u}+\frac{f_{,v}}{r}\right).
\end{eqnarray}

Also, we can obtain the energy-momentum tensors for the Brans-Dicke field part and the scalar field part:
\begin{eqnarray}
\label{eq:TBDuu}T^{\mathrm{BD}}_{uu} &=& \frac{1}{8 \pi \phi} (w_{,u} - 2hw) + \frac{\omega}{8 \pi \phi^{2}} w^{2},\\
\label{eq:TBDuv}T^{\mathrm{BD}}_{uv} &=& - \frac{z_{,u}}{8 \pi \phi} - \frac{gw+fz}{4 \pi r \phi},\\
\label{eq:TBDvv}T^{\mathrm{BD}}_{vv} &=& \frac{1}{8 \pi \phi} (z_{,v} - 2dz) + \frac{\omega}{8 \pi \phi^{2}} z^{2},\\
\label{eq:TBDthth}T^{\mathrm{BD}}_{\theta\theta} &=& \frac{r^{2}}{2 \pi \alpha^{2} \phi} z_{,u} + \frac{r}{4 \pi \alpha^{2} \phi} (gw+fz) + \frac{\omega}{4\pi \phi^{2}} \frac{r^{2}}{\alpha^{2}}wz,
\end{eqnarray}
\begin{eqnarray}
\label{eq:TSuu}\frac{T^{\Phi}_{uu}}{\phi} &=& \frac{1}{4 \pi \phi}W^{2},\\
\label{eq:TSuv}\frac{T^{\Phi}_{uv}}{\phi} &=& \frac{\alpha^{2}}{2\phi}V(S),\\
\label{eq:TSvv}\frac{T^{\Phi}_{vv}}{\phi} &=& \frac{1}{4 \pi \phi}Z^{2},\\
\label{eq:TSthth}\frac{T^{\Phi}_{\theta\theta}}{\phi} &=& \frac{r^{2}}{2\pi\alpha^{2} \phi} WZ - \frac{r^{2}}{\phi}V(S).
\end{eqnarray}

To implement double-null formalism into the numerical scheme, it is convenient to represent all equations as first order differential equations.
Note that
\begin{eqnarray}\label{eq:T_trace}
T^{\Phi} = - \frac{4}{\alpha^{2}}T^{\Phi}_{uv} + \frac{2}{r^{2}}T^{\Phi}_{\theta \theta}.
\end{eqnarray}
The Einstein equations for $\alpha_{,uv}$, $r_{,uv}$, and the field equation for $\phi$ are then coupled:
\begin{eqnarray}\label{eq:coupled}
\left( \begin{array}{ccc}
1 & 1/r & 1/\phi \\
0 & 1 & r/2\phi \\
0 & 0 & r
\end{array} \right)
\left( \begin{array}{c}
d_{,u} \\
f_{,v} \\
z_{,u}
\end{array} \right)
= \left( \begin{array}{c}
\mathfrak{A} \\
\mathfrak{B} \\
\mathfrak{C}
\end{array} \right)
\end{eqnarray}
where
\begin{eqnarray}
\label{eq:A}\mathfrak{A} &\equiv& -\frac{2\pi \alpha^{2}}{r^{2}\phi}T^{\Phi}_{\theta\theta} - \frac{1}{2r}\frac{1}{\phi}(gw+fz) -\frac{\omega}{2\phi^{2}}wz, \\
\label{eq:B}\mathfrak{B} &\equiv& - \frac{\alpha^{2}}{4r} - \frac{fg}{r} + \frac{4 \pi r}{\phi}T^{\Phi}_{uv} - \frac{1}{\phi}(gw+fz), \\
\label{eq:C}\mathfrak{C} &\equiv& - fz - gw - \frac{4\pi r}{3+2\omega} \left(\frac{WZ}{2 \pi} - 2 \alpha^{2}V \right).
\end{eqnarray}

After solving these coupled equations, we can write all equations:
\begin{eqnarray}
\label{eq:E1}f_{,u} &=& 2fh - \frac{r}{2 \phi} (w_{,u}-2hw) - \frac{r}{\phi}W^{2} - \frac{r \omega}{2 \phi^{2}} w^{2},\\
\label{eq:E2}g_{,v} &=& 2gd - \frac{r}{2 \phi} (z_{,v}-2dz) - \frac{r}{\phi}Z^{2} - \frac{r \omega}{2 \phi^{2}} z^{2},\\
\label{eq:E3}d_{,u} = h_{,v} &=& \frac{fg}{r^{2}} + \frac{\alpha^{2}}{4r^{2}} + \frac{gw+fz}{r\phi} - \frac{\omega}{2\phi^{2}}wz + \frac{2 \pi}{(3+2\omega)\phi} \left( \frac{WZ}{2\pi} - 2\alpha^{2} V \right) - \frac{WZ}{\phi},\\
\label{eq:E4}g_{,u} = f_{,v} &=& -\frac{fg}{r}-\frac{\alpha^{2}}{4r} - \frac{gw+fz}{2\phi} + \frac{2 \pi r}{(3+2\omega) \phi} \left( \frac{WZ}{2 \pi} - 2 \alpha^{2} V \right) + \frac{2 \pi r \alpha^{2}}{\phi}V,\\
\label{eq:s}z_{,u} = w_{,v} &=& -\frac{1}{r} \left( gw + fz + \frac{4\pi r}{3+2\omega} \left( \frac{WZ}{2\pi}-2 \alpha^{2}V \right) \right),
\end{eqnarray}
including the scalar field equation
\begin{eqnarray}
\label{eq:S}Z_{,u} = W_{,v} = -\frac{fZ}{r}-\frac{gW}{r}-\pi \alpha^{2}V'(S).
\end{eqnarray}

Now, equations of $\alpha_{,uv}$, $r_{,uv}$, $\phi_{,uv}$, and $S_{,uv}$ parts can be represented by first order differential equations.
We can then implement the same integration scheme used in previous papers \cite{Yeom1}\cite{Yeom2} to solve the Brans-Dicke theory.
We used the second order Runge-Kutta method \cite{nr}. Tests of the convergence are provided in Appendix~B.

\subsection{\label{sec:ini}Initial conditions and free parameters}

We need initial conditions for all functions ($\alpha, h, d, r, f, g, \phi, w, z, S, W, Z$) on the initial $u=u_{\mathrm{i}}$ and $v=v_{\mathrm{i}}$ surfaces, where we set $u_{\mathrm{i}}=v_{\mathrm{i}}=0$.

We have gauge freedom to choose the initial $r$ function. Although all constant $u$ and $v$ lines are null, there remains freedom to choose the distances between these null lines. Here, we choose $r(0,0)=r_{0}$, $f(u,0)=r_{u0}$, and $g(0,v)=r_{v0}$, where $r_{u0}<0$ and $r_{v0}>0$ such that the radial function for an in-going observer decreases and that for an out-going observer increases.

First, we assume that the gravitation constant $G=1/\phi$ is asymptotically $1$. Then, $\phi(u,0)=\phi(0,v)=1$ and $w(u,0)=z(0,v)=0$.

We use a shell-shaped scalar field, and hence its interior is not affected by the shell.
Thus, we can simply choose $S(u,0)=A$ and $\alpha(u,0)=1$. Also, $W(u,0)=h(u,0)=0$ holds.
Then, since the asymptotic mass function \cite{Waugh:1986jh}
\begin{eqnarray}
m(u,v) \equiv \frac{r}{2}\left( 1+4\frac{r_{,u}r_{,v}}{\alpha^{2}} \right),
\end{eqnarray}
should vanish at $u=v=0$, it is convenient to choose $r_{u0}=-1/2$ and $r_{v0}=1/2$.

We need more information to determine $d, g, z$, and $Z$ on the $v=0$ surface. We obtain $d$ from Equation~(\ref{eq:E3}), $g$ from Equation~(\ref{eq:E4}), $z$ from Equation~(\ref{eq:s}), and $Z$ from Equation~(\ref{eq:S}), respectively.

We can choose an arbitrary function for $S(0,v)$ to induce a collapsing pulse. In this paper, we use
\begin{eqnarray}
S(0,v)=A \left(1-\frac{v}{v_{\mathrm{f}}}\right) + \frac{A}{2 \pi} \sin \left( 2 \pi \frac{v}{v_{\mathrm{f}}} \right)
\end{eqnarray}
for $0\leq v \leq v_{\mathrm{f}}$ and otherwise $S(0,v)=0$, where $v_{\mathrm{f}}$ is the width of the pulse and $A$ is the amplitude.
Then we obtain
\begin{eqnarray}
Z(0,v)= -\frac{2A}{v_{\mathrm{f}}} \sin^{2} \left(\pi \frac{v}{v_{\mathrm{f}}} \right)
\end{eqnarray}
for $0\leq v \leq v_{\mathrm{f}}$ and otherwise $Z(0,v)=0$.
This implements one pulse of energy ($T_{vv} \sim Z^{2}$) along the out-going null direction by a differentiable function $Z(0,v)$.

Also, from Equation~(\ref{eq:E2}), we can use $d = rZ^{2}/2g\phi$ on the $u=0$ surface and we obtain $d(0,v)$. By integrating $d$ along $v$, we have $\alpha(0,v)$.

We need more information for $h, f, w,$ and $W$ on the $u=0$ surface. We obtain $h$ from Equation~(\ref{eq:E3}), $f$ from Equation~(\ref{eq:E4}), $w$ from Equation~(\ref{eq:s}), and $W$ from Equation~(\ref{eq:S}), respectively. This finishes the assignments of initial conditions.

We choose $r_{0}=10$, leaving the three free parameters $(\omega, A, v_{\mathrm{f}})$, where $\omega$ is the Brans-Dicke coupling parameter, $A$ is the amplitude of a pulse of the scalar field, and $v_{\mathrm{f}}$ is the width of the pulse.
Here, we assume that there is no potential term in the matter field side.

\section{\label{sec:res}Responses of the Brans-Dicke field due to gravitational collapses}

\subsection{\label{sec:sim}Simulation results}

\begin{figure}
\begin{center}
\includegraphics[scale=0.3]{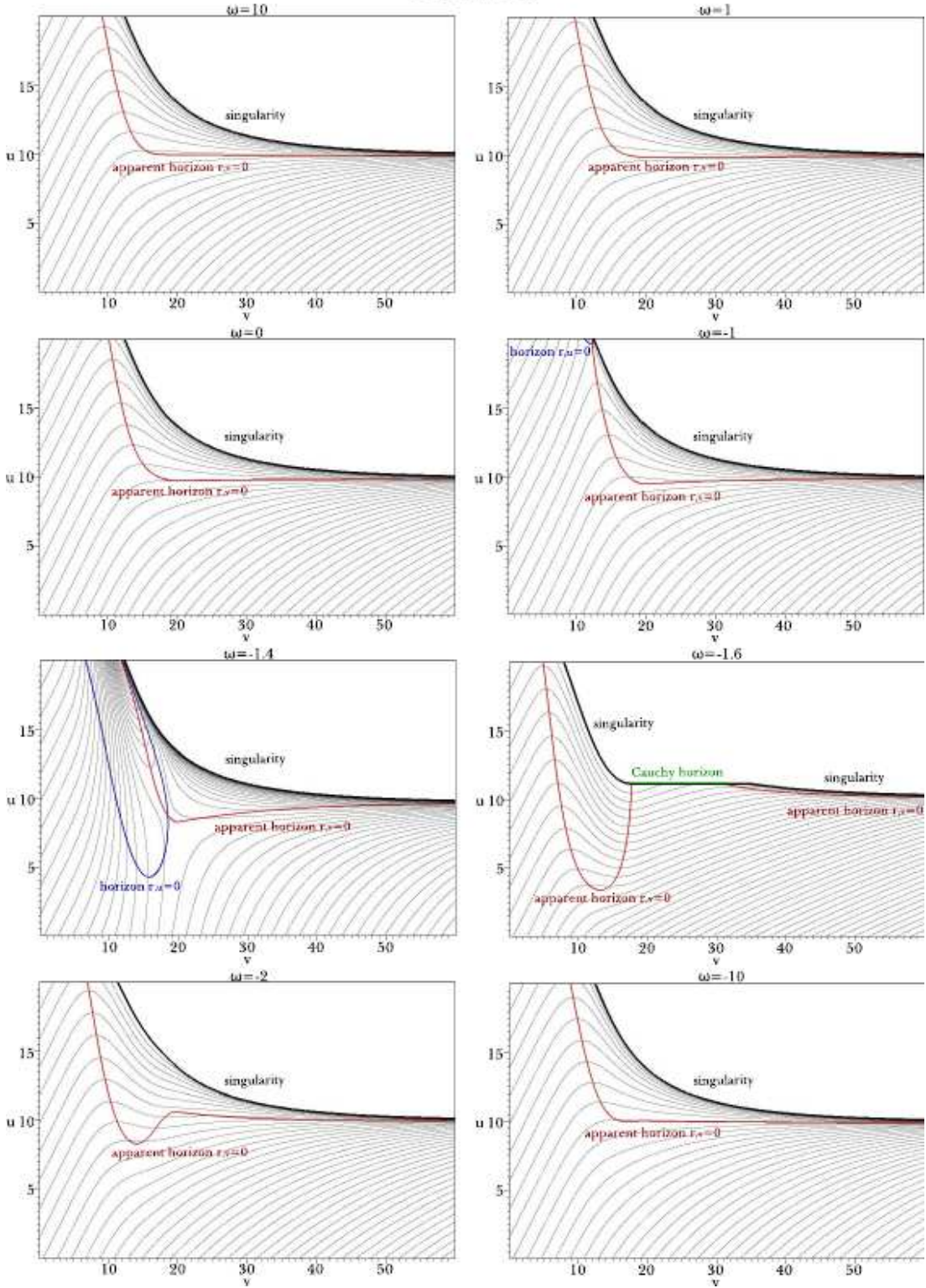}
\caption{\label{fig:varying_omega}Contour diagrams of the radial function $r$. $A=0.75$, $v_{\mathrm{f}}=20$, and $\omega = 10, 1, 0, -1, -1.4, -1.6, -2, -10$. Spacing of each contour is $1$. Here, we plotted singularities, horizons ($r_{,v}=0$ and $r_{,u}=0$), and Cauchy horizons.}
\end{center}
\end{figure}

We report the simulation results in Figure~\ref{fig:varying_omega}.
We fixed $A=0.75$ and $v_{\mathrm{f}}=20$, and varied $\omega = 10,1,0,-1,-1.4,-1.6,-2,-10$.
Note that if $\omega=-1.5$, because of Equation~(\ref{eq:phi}), the equation becomes singular.

Figure~\ref{fig:varying_omega} contains contour diagrams of the radial function $r$ for each parameters.
We plotted singularities, trapping (apparent) horizons, and Cauchy horizons.
\begin{itemize}
\item If the radial function becomes $0$, since the $r=0$ curve is space-like and all equations become singular, it is reasonable to interpret this region as a singularity.

\item To define a black hole using local geometry, we may use apparent horizons \cite{Hawking:1973uf}\cite{Wald:1984rg} or trapping horizons \cite{Ashtekar:2004cn} for an out-going observer, i.e., $r_{,v}=0$.
If there is inflation at some point in space-time, the inflating region can be defined by an in-going observer, i.e., $r_{,u}=0$, so that the in-going observer sees an increase of the area function.
We plotted two horizons: $r_{,v}=0$ and $r_{,u}=0$.

\item If an apparent horizon contracts to a singularity, we can find a null cutoff line where we cannot calculate,
because we do not have sufficient information to determine this region via the singularity;
hence, we can identify the region as a Cauchy horizon \cite{Hawking:1973uf}\cite{Wald:1984rg}.
\end{itemize}

\subsection{\label{sec:obs}Observations}

\begin{figure}
\begin{center}
\includegraphics[scale=0.75]{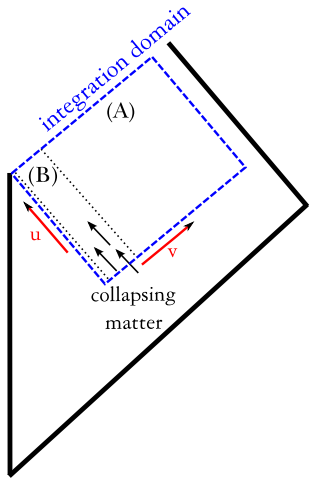}
\caption{\label{fig:domain}The integration domain of the double-null simulation. Region (A) is for the late time behavior; (B) is for the dynamical behavior.}
\end{center}
\end{figure}

In this subsection, we list some interesting observations for Figure~\ref{fig:varying_omega}. In the next section, we interpret the observations.

\begin{enumerate}
\item Eventually a black hole is formed, i.e., a space-like singularity and an apparent horizon are observed.

\item If $|\omega|$ is sufficiently large, the causal structures become similar to each other and the apparent horizons are always space-like. See $\omega=10$ and $-10$ cases.

\item As $\omega$ decreases, if $\omega>-1.5$, the apparent horizon approaches the event horizon in the time-like direction. See $\omega=1, 0, -1, -1.4$.

\item If $\omega<-1.5$ and $\omega$ is sufficiently small, the apparent horizon oscillates and eventually approaches the event horizon in the space-like direction. See $\omega=-1.6, -2$.

\item If $\omega \gtrsim -1.5$, the causal structure can contain an instantaneously inflating region, i.e., there can exist an $r_{,u}=0$ horizon.
As $\omega$ approaches $-1.5$, the $r_{,u}=0$ horizon appears from upper $u$ to lower $u$. Compare $\omega=-1$ and $\omega=-1.6$.

\item If $\omega \lesssim -1.5$, the apparent horizon can shrink to a singularity and form a Cauchy horizon. However, eventually a black hole will be formed. See $\omega=-1.6$.
\end{enumerate}

Observations $1, 2, 3,$ and $4$ pertain to asymptotic behaviors after the collapse of matter. We call them \textit{late time behaviors}.
Observations $5$ and $6$ pertain to fully dynamical behaviors during the gravitational collapse. We call them \textit{dynamical behaviors} (Figure~\ref{fig:domain}).
We discuss and interpret the late time behaviors and dynamical behaviors in the following section.

\section{\label{sec:inter}Interpretations}

\subsection{\label{sec:lat}Late time behaviors and dynamics of the Brans-Dicke field}

From the work of previous authors using Brans-Dicke theory, we know that a gravitational collapse should lead to the emergence of an Einstein black hole asymptotically \cite{Hawking:1972qk}.
As the black holes emerge to Einstein black holes, some interesting observations can be made.
To understand these observations, in this section, we discuss the large $|\omega|$ case and the $\omega \sim -1.5$ case independently.

\begin{figure}
\begin{center}
\includegraphics[scale=0.3]{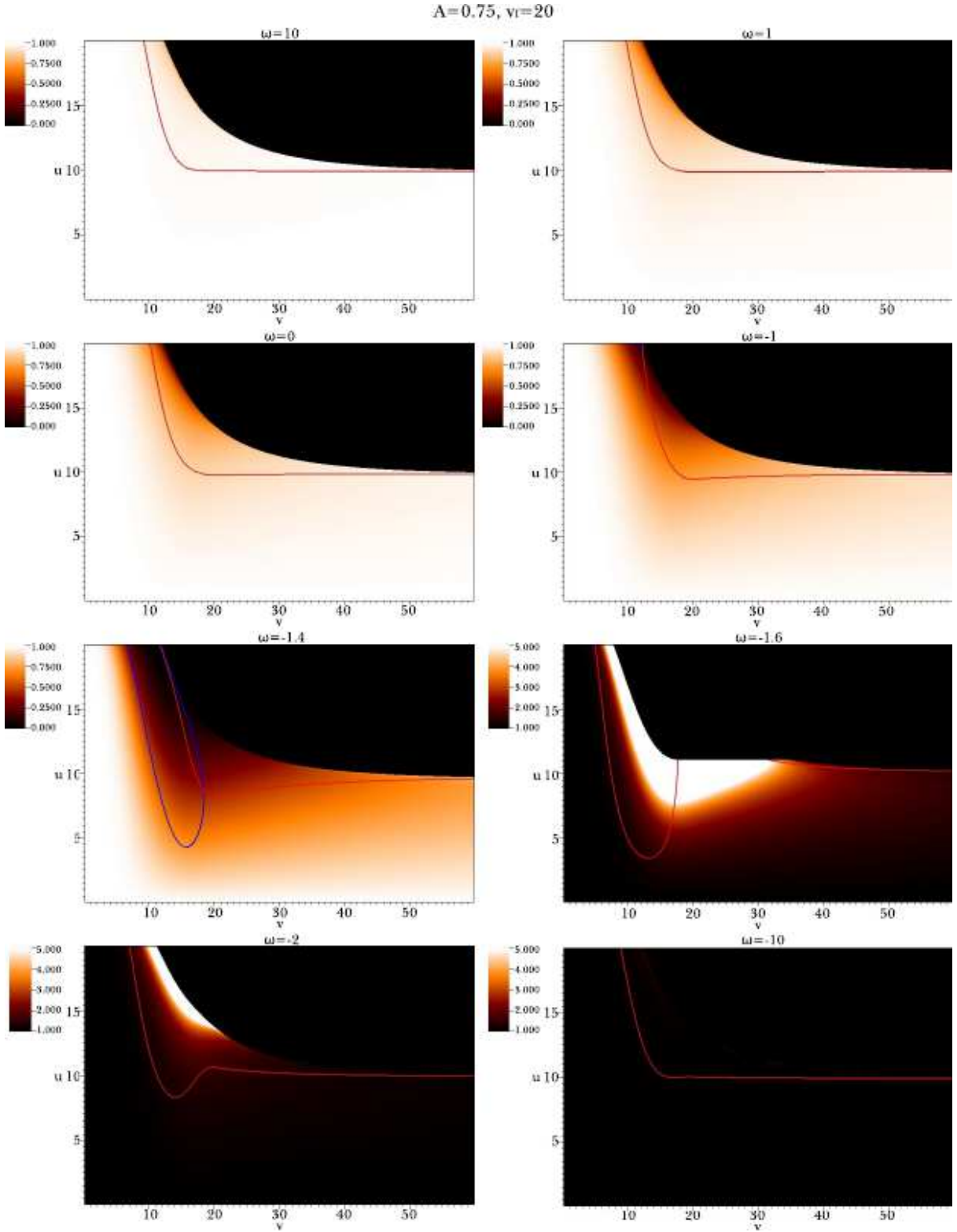}
\caption{\label{fig:Sb}Plot of the Brans-Dicke field $\phi$. $A=0.75$, $v_{\mathrm{f}}=20$, and $\omega = 10, 1, 0, -1, -1.4, -1.6, -2, -10$.
If the color changes, it shows that the Brans-Dicke field has dynamical behaviors. If $\omega > -1.5$, then $\phi$ decreases during the matter collapse;
if $\omega < -1.5$, then $\phi$ increases during the matter collapse.}
\end{center}
\end{figure}

\subsubsection{The large $|\omega|$ limit}

If $|\omega|$ is sufficiently large, the field equation of the Brans-Dicke field (Equation~(\ref{eq:phi})) becomes a free scalar field equation.
Then, if $\phi$ was asymptotically $1$ and initially had no dynamics, it will not be affected by gravitation, since the field equation is a free scalar field equation.
(See the cases of $\omega=10$ and $\omega=-10$ in Figure~\ref{fig:Sb}.)
Therefore, the dynamics during the collapse of matter will be similar to that of Einstein gravity and we can ignore the effect of the Brans-Dicke field.
This explains why the black holes in the large $|\omega|$ limit resemble Einstein black holes.

Note that, in these cases, space-like apparent horizons are observed.
If there is no supply of matter, the horizon should be null.
However, during the matter collapse, there is some scattered energy along the out-going direction,
and part of this energy will come to the black hole later.
This explains why the horizons are still space-like and not null after $v_{\mathrm{f}}=20$.

\begin{figure}
\begin{center}
\includegraphics[scale=0.3]{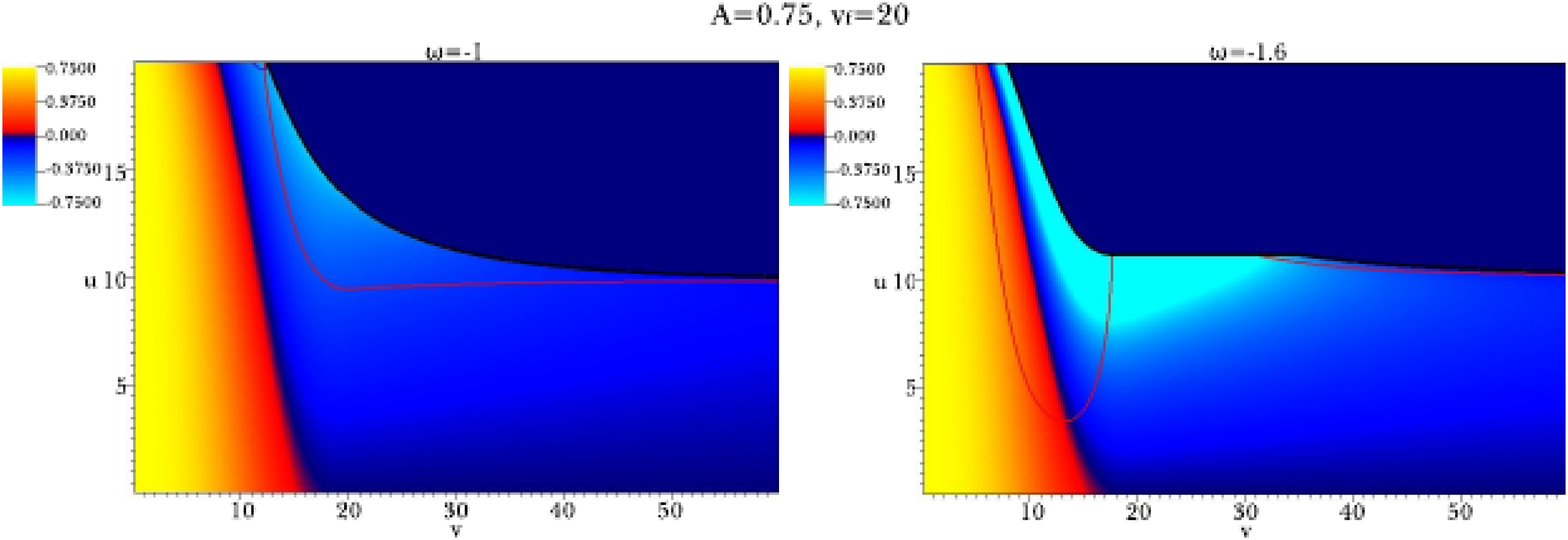}
\caption{\label{fig:Sm}Plot of the scalar field $S$. $A=0.75$, $v_{\mathrm{f}}=20$, and $\omega = -1$ and $\omega = -1.6$. These figures show that,
during collapse of matter, $S$ decreases for both the in-going and out-going directions. Therefore, $S_{,u}S_{,v}\geq0$ holds during matter supply.
For other parameters, the behaviors are similar during the matter collapse.}
\end{center}
\end{figure}

\subsubsection{$\omega$ near $-1.5$}

If $\omega$ is near $-1.5$, we will see dynamical effects of the Brans-Dicke field.
Figure~\ref{fig:Sb} shows that if $\omega$ is greater than $-1.5$, it decreases during the matter collapse; if $\omega$ is less than $-1.5$, it increases during the matter collapse.
We discuss and interpret these phenomena below.

We have the equation for the Brans-Dicke field:
\begin{eqnarray}
\phi_{,uv} &=& -\frac{1}{r} \left( r_{,v}\phi_{,u} + r_{,u}\phi_{,v} + \frac{2r}{3+2\omega} S_{,u}S_{,v} \right).
\end{eqnarray}
During the matter collapse, the $S_{,u}S_{,v}$ term is dominant.
Also, in our initial condition, $S_{,u}S_{,v}$ is always positive during the matter collapse (Figure~\ref{fig:Sm}).
Then, the total sign of $\phi_{,uv}$ is determined by the sign of $3+2\omega$.
Therefore, if $\omega > -1.5$, the total sign of $\phi_{,uv}$ becomes negative, and hence the Brans-Dicke field tends to decrease.
If $\omega < -1.5$, by the same reasoning, the Brans-Dicke field tends to increase.

After the matter supply ends, the Brans-Dicke field will be relaxed to the asymptotic value, since the black hole should approach an Einstein black hole.
Hence, if $\omega > -1.5$, the decreased Brans-Dicke field tends to increase for an out-going observer around the horizon.
Note that the area function in the Einstein frame is $\phi \mathcal{A}$ \cite{Kang:1996rj}, where $\mathcal{A}$ is the area of a black hole in the Jordan frame.
We know that, according to the area theorem, the area of the Einstein frame ($\phi \mathcal{A}$) will always increase and be space-like.
Then,
\begin{eqnarray}
\delta(\phi \mathcal{A}) \geq 0.
\end{eqnarray}
If there is no collapse of additional matter, $\delta (\phi \mathcal{A}) \sim 0$ and $\phi \mathcal{A} \sim \mathrm{const}$.
Therefore, if $\phi$ increases along the horizon, the area should shrink and be time-like.
If $\omega < -1.5$, by the same reasoning, we can explain that the horizon contracts along a space-like direction.

\subsection{\label{sec:dyn}Dynamical behaviors and energy conditions}

Now we turn to a discussion of fully dynamical behaviors during the matter collapse.
We have observed interesting behaviors of $r_{,v}=0$ or $r_{,u}=0$ horizons.
These behaviors should be consistent with energy-momentum tensors.
Therefore, in this section, we look at details of the energy-momentum tensors and energy conditions
that cause the dynamical behaviors.
We discuss two different cases, $\omega \gtrsim -1.5$ and $\omega \lesssim -1.5$.

To understand the dynamical behaviors, we plot the energy-momentum tensor components (Figure~\ref{fig:lessT}).
For any observer whose four velocity is $n^{\mu}$, the observer feels the energy density by
\begin{eqnarray}
T_{\mu\nu}n^{\mu}n^{\nu}.
\end{eqnarray}
Therefore, if an observer moves along an in-going null direction, the observer feels the energy density by $T_{uu}$;
if an observer moves along an out-going null direction, the observer feels the energy density by $T_{vv}$.
Note that the null energy condition is violated if $T_{uu}$ or $T_{vv}$ has a negative region.

Around horizons ($r_{,u}=0$ or $r_{,v}=0$), we can simplify Equations~(\ref{eq:E1}) and (\ref{eq:E2}) by
\begin{eqnarray}
r_{,uu} &=& -4\pi rT_{uu},\\
r_{,vv} &=& -4\pi rT_{vv}.
\end{eqnarray}
We know that asymptotically $r_{,u}<0$ and $r_{,v}>0$; i.e., initially,
an in-going observer sees a decrease of area and an out-going observer sees an increase of area.
Using these observations, we remark on two cases where the null energy condition should be violated.
\begin{itemize}
\item We can define the outer part of an $r_{,u}=0$ horizon if the sign of $r_{,u}$ changes from $-$ to $+$ along an in-going null direction;
the inner part of an $r_{,u}=0$ changes its sign from $+$ to $-$ along an in-going null direction.
Then, around the outer $r_{,u}=0$ horizon, $r_{,uu}$ should be positive, and hence $T_{uu}<0$.
\item If $r_{,vv}<0$ around the (outer) $r_{,v}=0$ horizon, the horizon is space-like;
if $r_{,vv}>0$ around the $r_{,v}=0$ horizon, the horizon is time-like.
Therefore, to see a time-like $r_{,v}=0$ horizon, $T_{vv}<0$ is required.
\end{itemize}

Note that, $T_{uv}$ component is related to the vacuum energy or cosmological constant term.
Therefore, if $T_{uv}>0$, then it may cause an increase of the area; if $T_{uv}<0$, then it may cause a decrease of the area.
We can write Equation~(\ref{eq:E4}) around the $r_{,u}=0$ horizon as
\begin{eqnarray}
r_{,uv} = -\frac{\alpha^2}{4r}+4\pi rT_{uv}.
\end{eqnarray}
If there is inflation, an out-going null observer sees an increase of area.
This requires that $r_{,uv}>0$, and hence $T_{uv}$ should be sufficiently large:
\begin{eqnarray}
\frac{\alpha^2}{16 \pi r^{2}} \leq T_{uv}.
\end{eqnarray}
If the inflation tends to end, the opposite situation should occur and $T_{uv}$ should be sufficiently small so that $r_{,uv}<0$.
Therefore, we need to check the sign of $T_{uv}$ components if there is an $r_{,u}=0$ horizon.

\subsubsection{$\omega \gtrsim -1.5$}

\begin{figure}
\begin{center}
\includegraphics[scale=0.25]{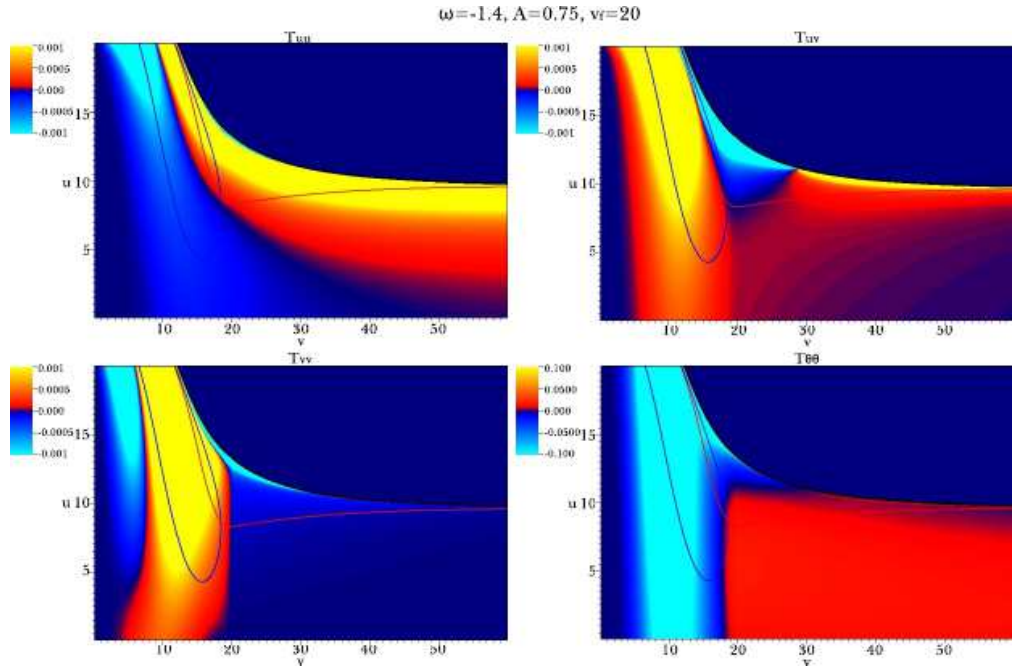}
\caption{\label{fig:lessT}The energy-momentum tensors for ($\omega=-1.4$, $A=0.75$, $v_{\mathrm{f}}=20$).}
\end{center}
\end{figure}
\begin{figure}
\begin{center}
\includegraphics[scale=0.25]{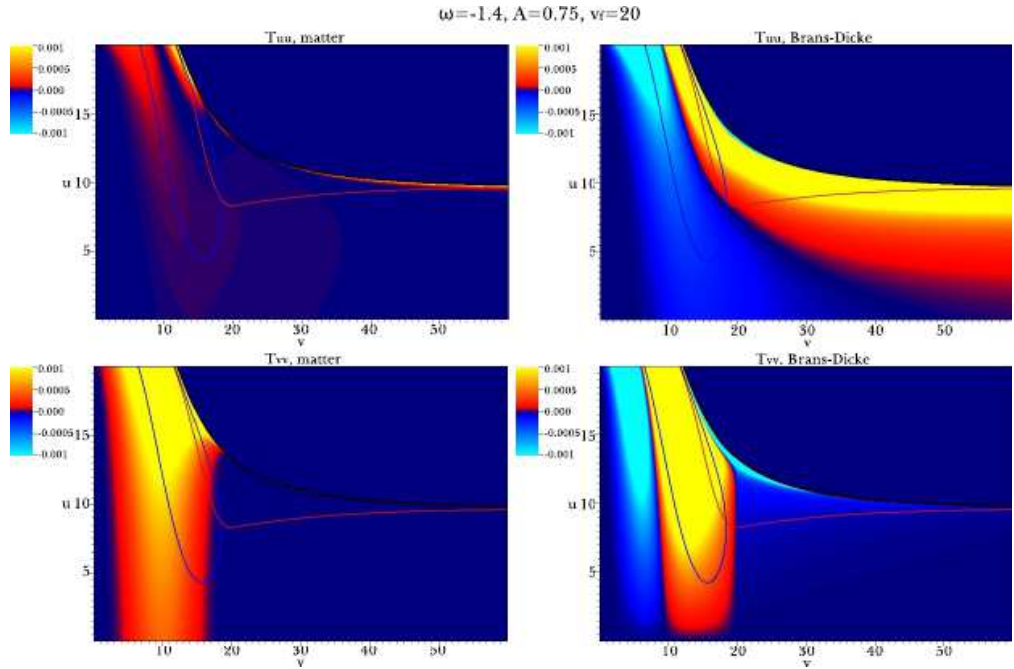}
\caption{\label{fig:lessT2}The energy-momentum tensors of the matter sectors ($T^{\Phi}_{uu}/\phi$, $T^{\Phi}_{vv}/\phi$) and the Brans-Dicke sectors ($T^{\mathrm{BD}}_{uu}$, $T^{\mathrm{BD}}_{vv}$) for ($\omega=-1.4$, $A=0.75$, $v_{\mathrm{f}}=20$).}
\end{center}
\end{figure}

\begin{figure}
\begin{center}
\includegraphics[scale=0.25]{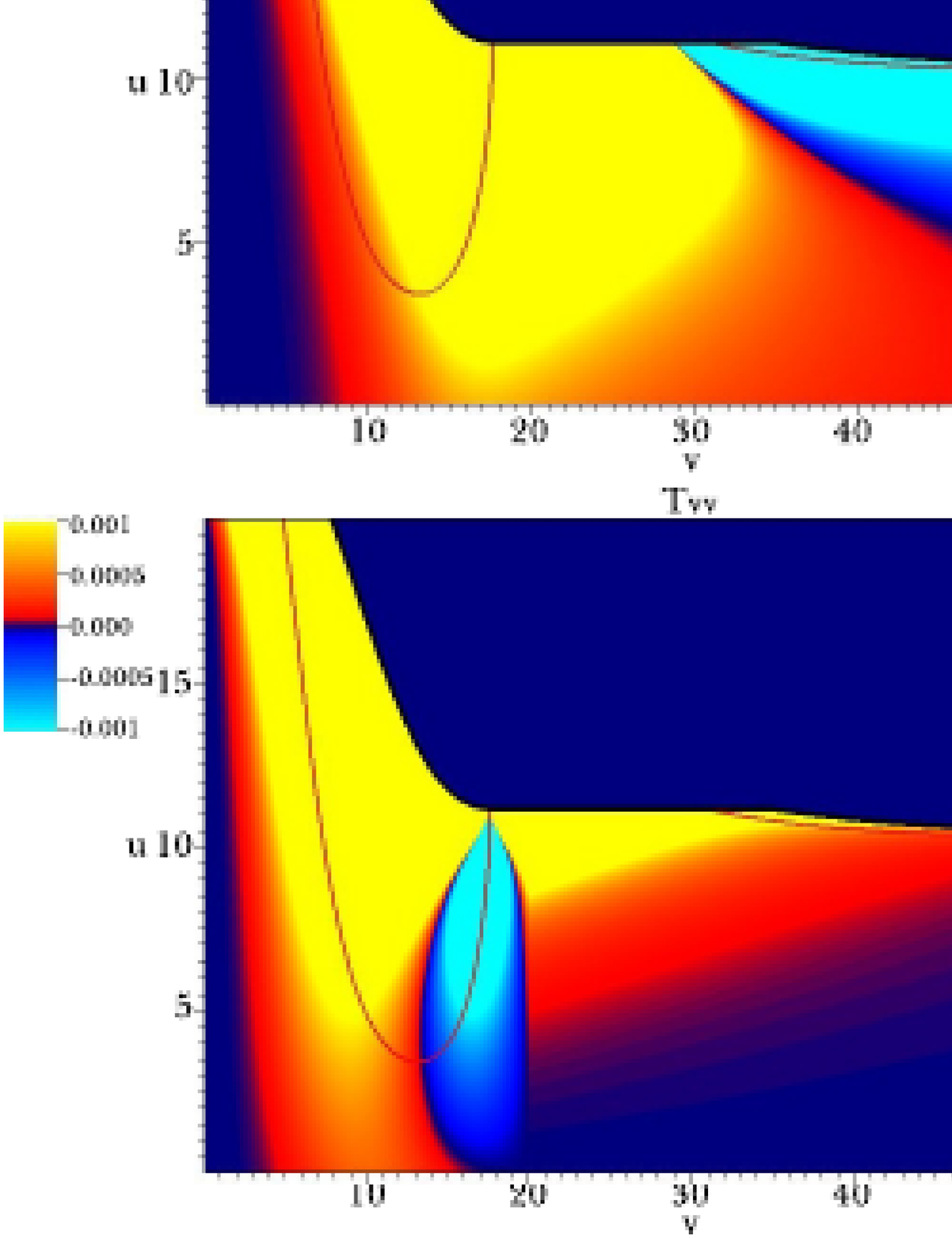}
\caption{\label{fig:moreT}The energy-momentum tensors for ($\omega=-1.6$, $A=0.75$, $v_{\mathrm{f}}=20$).}
\end{center}
\end{figure}
\begin{figure}
\begin{center}
\includegraphics[scale=0.25]{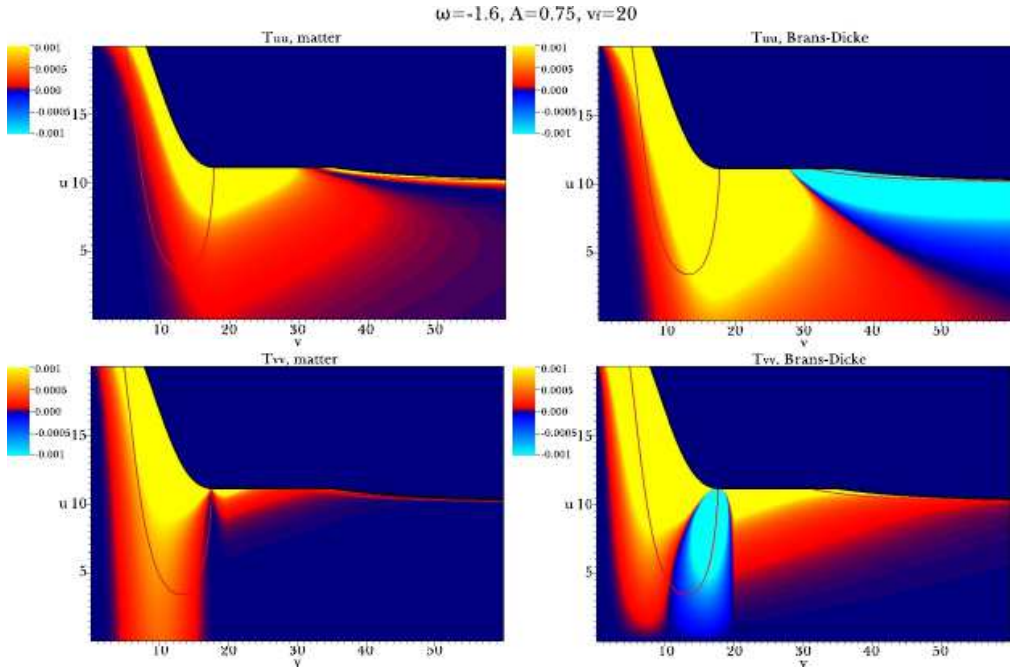}
\caption{\label{fig:moreT2}The energy-momentum tensors of the matter sectors ($T^{\Phi}_{uu}/\phi$, $T^{\Phi}_{vv}/\phi$)
and the Brans-Dicke sectors ($T^{\mathrm{BD}}_{uu}$, $T^{\mathrm{BD}}_{vv}$) for ($\omega=-1.6$, $A=0.75$, $v_{\mathrm{f}}=20$).}
\end{center}
\end{figure}

In Figure~\ref{fig:lessT}, we can see the energy-momentum tensor components of the $\omega \gtrsim -1.5$ case.
The outer $r_{,u}=0$ horizon was possible since the $T_{uu}<0$ condition holds.
Also, $r_{,v}=0$ could bend in the time-like direction since the $T_{vv}<0$ condition holds.
Around the $r_{,u}=0$ horizon, the $T_{uv}$ component changes its sign.
We can interpret that, as the area increases, some effective vacuum energy can be gained from the Brans-Dicke field.
However, as the effective vacuum energy changes its sign, the area ceases to increase and eventually decreases for an in-going observer.
We call this an \textit{instantaneous inflation} driven by gravitational collapse.

We can compare the energy-momentum tensors by two sectors:
\begin{eqnarray}
T_{\mu\nu} = \frac{T^{\Phi}_{\mu\nu}}{\phi} + T^{\mathrm{BD}}_{\mu\nu},
\end{eqnarray}
where $T^{\Phi}_{\mu\nu}$ and $T^{\mathrm{BD}}_{\mu\nu}$ are defined by Equations~\ref{eq:T_scalar} and \ref{eq:T_BD}.
In Figure~\ref{fig:lessT2}, we compare these two sectors.
During the matter supply, the matter part and the Brans-Dicke part are equally important for $T_{vv}$.
However, in other cases, the Brans-Dicke part dominates.

\subsubsection{$\omega \lesssim -1.5$}

In Figure~\ref{fig:moreT}, we can see the energy-momentum tensor components of the $\omega \lesssim -1.5$ case.
One interesting observation is that $T_{vv}$ has a negative region when the $r_{,v}=0$ horizon shrinks.
The negative part of $T_{vv}$ originates from the negative part of $T^{\mathrm{BD}}_{vv}$ (Figure~\ref{fig:moreT2}).
As $v$ increases, the sign of $T^{\mathrm{BD}}_{vv}$ is changed by $+, -, +$.
This is an opposite behavior than the $\omega \gtrsim -1.5$ case; in this case, $T^{\mathrm{BD}}_{vv}$ was changed as $-, +, -$.

These differences are consistent with the late time behaviors.
When the matter collapse finishes, $T^{\mathrm{BD}}_{vv}$ dominates the total energy-momentum tensor.
Hence, in the $\omega \lesssim -1.5$ case, the apparent horizon is space-like;
in the $\omega \gtrsim -1.5$ case, the apparent horizon is time-like.

\subsubsection{Violation of weak cosmic censorship}

Finally, we discuss the violation of weak cosmic censorship \cite{Wald:1984rg} for the $\omega \lesssim -1.5$ case.
The existence of a Cauchy horizon that is not hidden by an apparent horizon implies the violation of weak cosmic censorship.
However, it is interesting to check whether the existence of the Cauchy horizon can be maintained infinitely
or there should be a singularity and an apparent horizon, as in the ($\omega=-1.6$, $A=0.75$, $v_{\mathrm{f}}=20$) case.
We argue that if the asymptotic location of the singularity has an upper $u$ coordinate relative to the location of the Cauchy horizon,
we cannot see the singularity forever.

We briefly remark on a property of our double-null coordinates.
\begin{itemize}
\item Eventually, our black holes should approach a Schwarzschild black hole, and in this limit, we can use the Kruskal coordinates \cite{d'Inverno:1992rk} so that
\begin{eqnarray}
uv = (r-2M) \exp \left( \frac{r}{2M} \right),
\end{eqnarray}
where $M$ is the mass of the Schwarzschild black hole.
Then, the singularity is $uv=-2M$ and the horizon is $uv=0$.

If we choose $u=0$ at the horizon, then for a fixed $v$, the distance between the horizon and the singularity in terms of the $u$ coordinate becomes
$\Delta u \propto 1/v$.
Then, as $v$ increases, the singularity will approach the horizon in terms of the $u$ coordinate.

\item The location of the apparent horizon $r_{,v}=0$ will follow the case of a Schwarzschild black hole.
Also, we know that the horizon of a Schwarzschild black hole is only determined by the mass.
Therefore, if we fix $A$ and $v_{\mathrm{f}}$, and hence we fix the asymptotic mass, for any $\omega$,
we will see asymptotically the same $u$ coordinate for the apparent horizon.

\item Therefore, the final location of a singularity is the same if we fix $A$ and $v_{\mathrm{f}}$.
This is observed in Figure~\ref{fig:varying_omega}.
\end{itemize}

Therefore, first, to find the location of the singularity, for given $A$ and $v_{\mathrm{f}}$, we will see a black hole of the $\omega=-1.4$ case.
\footnote{Of course, we could compare other $\omega$ values. However, the $\omega=-1.4$ case clearly shows asymptotic time-like horizons.
The location of a singularity should have upper $u$ coordinate relative to that of a time-like horizon.
Therefore, the $\omega=-1.4$ case is useful to clarify the location of the singularity.}
Second, we compare the location of a Cauchy horizon for the $\omega=-1.6$ case with the same $A$ and same $v_{\mathrm{f}}$.
If the location of a Cauchy horizon in $\omega=-1.6$ has smaller $u$ than the location of a singularity in $\omega=-1.4$, this confirms that
the existence of a Cauchy horizon can be maintained infinitely.

\begin{figure}
\begin{center}
\includegraphics[scale=0.3]{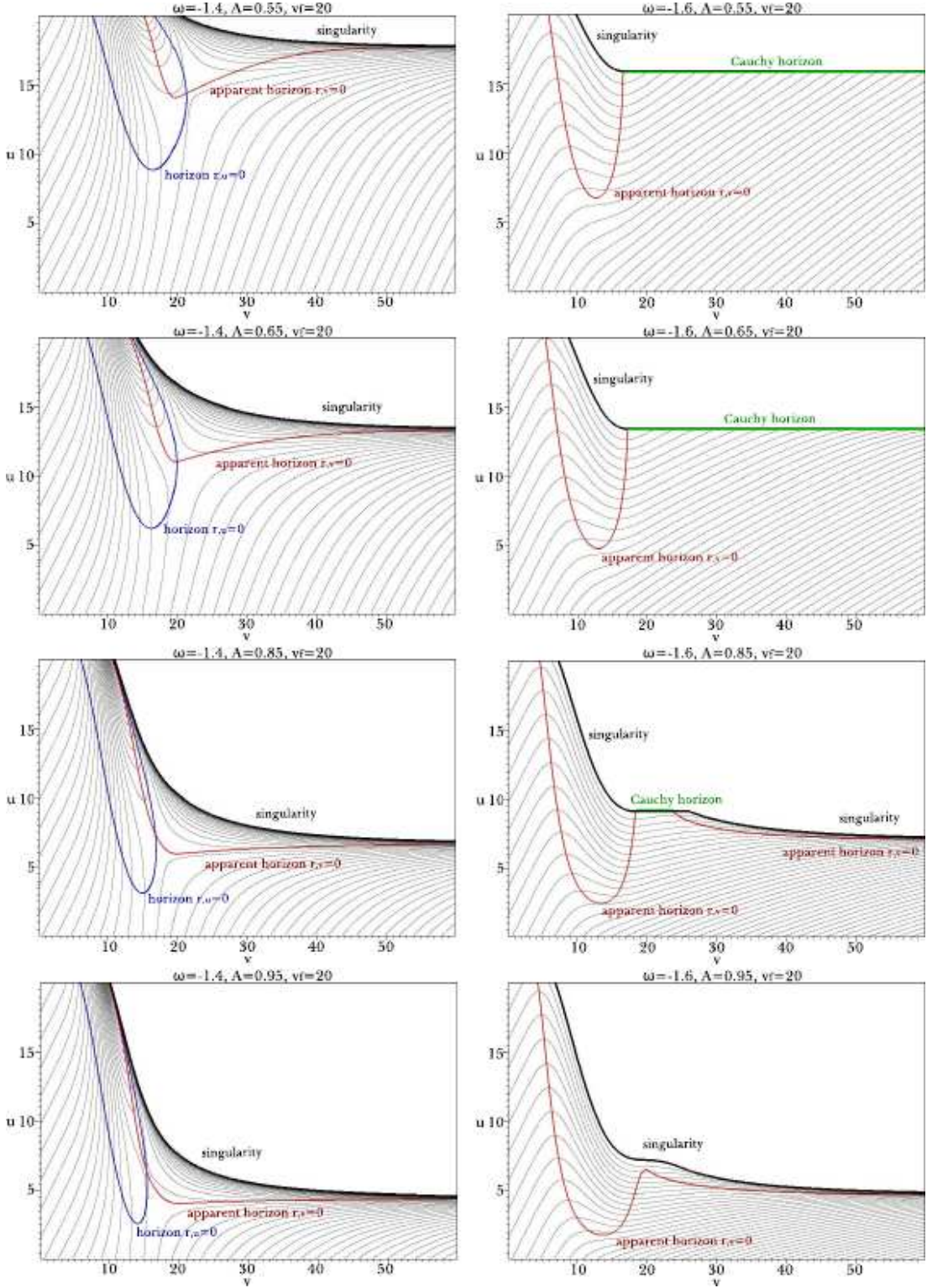}
\caption{\label{fig:varying_A}Contour diagrams of the radial function $r$. $v_{\mathrm{f}}=20$, $A = 0.55, 0.65, 0.85, 0.95$, and $\omega=-1.4$ and $-1.6$. Spacing is $1$.}
\end{center}
\end{figure}

\begin{figure}
\begin{center}
\includegraphics[scale=0.3]{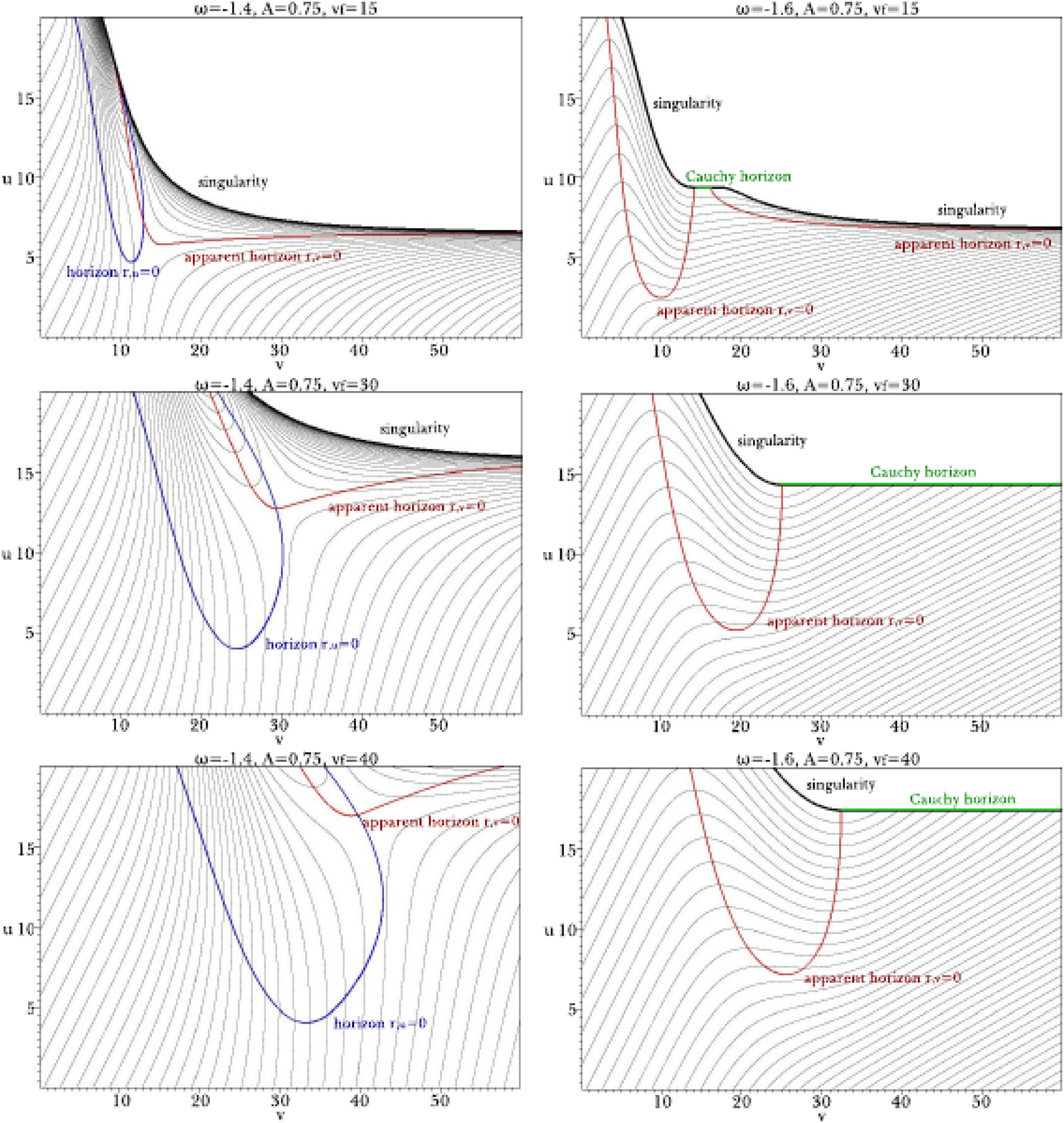}
\caption{\label{fig:varying_vf}Contour diagrams of the radial function $r$. $A=0.75$, $v_{\mathrm{f}}=15, 30, 40$, and $\omega = -1.4$ and $-1.6$. Spacing is $1$.}
\end{center}
\end{figure}

In Figures~\ref{fig:varying_A} and \ref{fig:varying_vf}, we present contour diagrams with various initial conditions.

In Figure~\ref{fig:varying_A}, we fixed $v_{\mathrm{f}}=20$ and changed $A=0.55, 0.65, 0.85, 0.95$. Also, we compared $\omega=-1.4$ and $\omega=-1.6$.
In the cases of $A=0.85$ and $0.95$, the locations of the singularity between $\omega=-1.4$ and $\omega=-1.6$ converge to the same location.
This confirms our previous remark on the double-null coordinates.
Therefore, in the cases of $A=0.55$ and $0.65$, we can check that the singularity will not appear below the Cauchy horizon,
and hence the Cauchy horizon will be maintained infinitely.

In Figure~\ref{fig:varying_vf}, we fixed $A=0.75$ and changed $v_{\mathrm{f}}=15, 30, 40$. Also, we compared $\omega=-1.4$ and $\omega=-1.6$.
As in the previous paragraph, we can see that $v_{\mathrm{f}}=30$ and $40$ cases allow infinite extension of the Cauchy horizon.

\section{\label{sec:con}Conclusion}

\begin{figure}
\begin{center}
\includegraphics[scale=0.75]{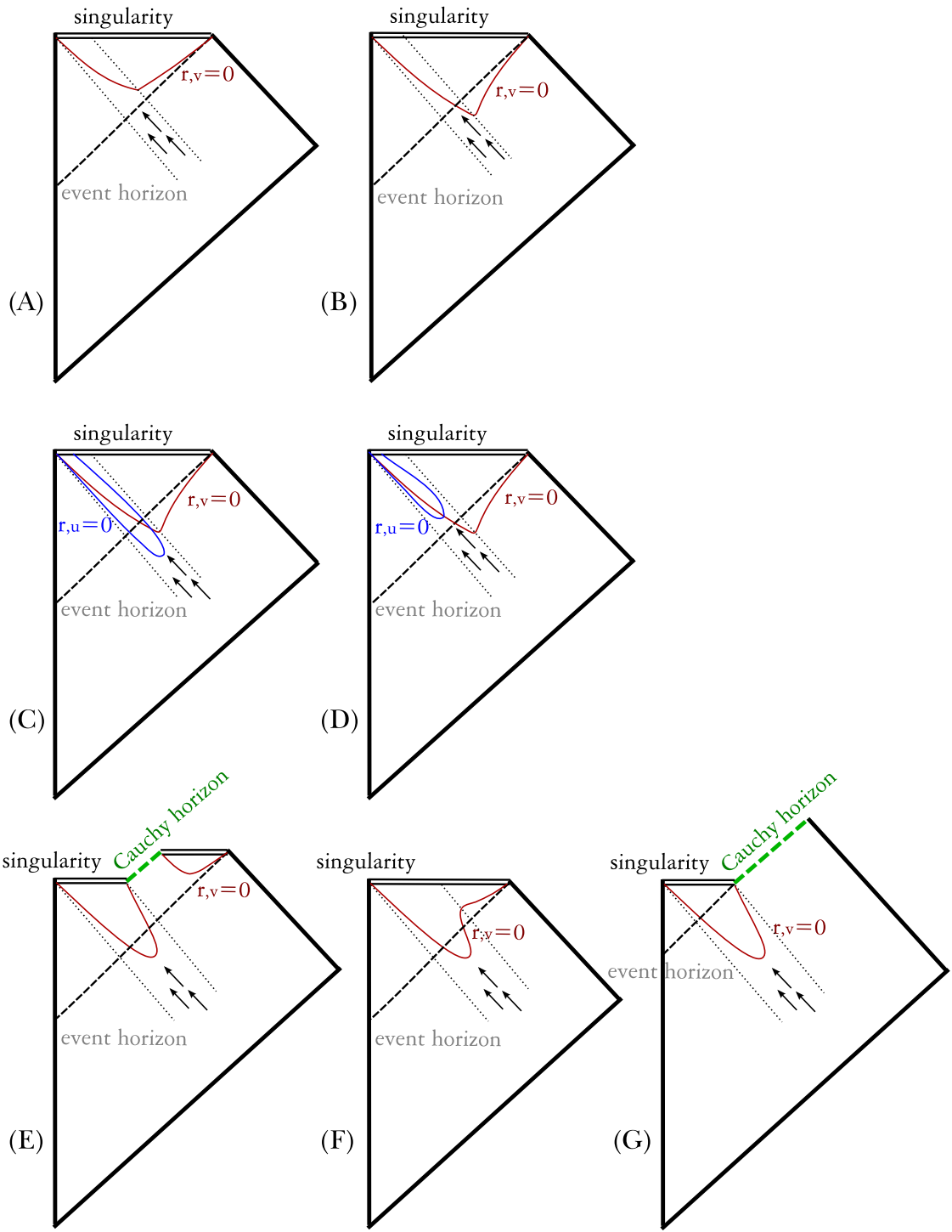}
\caption{\label{fig:classification}Classification of solutions.}
\end{center}
\end{figure}

\subsection{\label{sec:sum}Summary of results}

We study responses of the Brans-Dicke field due to gravitational collapses of scalar field pulses using numerical simulations.
In this paper, we used the double-null coordinates to implement the numerical simulations.

If the absolute value of the Brans-Dicke coupling constant $|\omega|$ is sufficiently large,
the responses of the Brans-Dicke field are negligible and eventually an Einstein black hole will be formed.
However, if $\omega$ is sufficiently small such that the Brans-Dicke field becomes sufficiently dynamic,
we can see fully dynamical back-reactions of the Brans-Dicke field.

If we supply a scalar field pulse, it will asymptotically form a black hole via dynamical interactions of the Brans-Dicke field.
Hence, we observed the responses of the Brans-Dicke field by two different regions (Figure~\ref{fig:domain}).
First, we observed the late time behaviors \textit{after} the gravitational collapse, which includes formations of a singularity and an apparent horizon.
Second, we observed the fully dynamical behaviors \textit{during} the gravitational collapse and viewed the energy-momentum tensor components.

For the late time behaviors, if $\omega$ is greater (or smaller) than $-1.5$,
the Brans-Dicke field decreases (or increases) during the gravitational collapse (Figure~\ref{fig:Sb}).
Since the Brans-Dicke field should be relaxed to the asymptotic value with the elapse of time,
the final apparent horizon becomes time-like (or space-like) (Figure~\ref{fig:varying_omega}).

For the dynamical behaviors, we observed the energy-momentum tensors around $\omega \sim -1.5$.
If $\omega \gtrsim -1.5$, the $T_{uu}$ component can be negative at the outside of the black hole.
This allows an instantaneous inflating region during the gravitational collapse (Figure~\ref{fig:lessT}).
If $\omega \lesssim -1.5$, the oscillation of the $T_{vv}$ component allows the apparent horizon to shrink (Figure~\ref{fig:moreT}).
In this case, we could find a combination of initial conditions that allows a violation of weak cosmic censorship (Figures~\ref{fig:varying_A} and \ref{fig:varying_vf}).

In Figure~\ref{fig:classification}, we summarize and classify all solutions in this paper.
\begin{itemize}
\item[(A)] If $|\omega|$ is sufficiently large, the $r_{,v}=0$ horizon is space-like.
\item[(B)] If $\omega>-1.5$ is sufficiently small, then the $r_{,v}=0$ horizon can be time-like after the matter collapse.
\item[(C), (D)] If $\omega \gtrsim -1.5$, there can be $r_{,u}=0$ horizons. After the matter collapse, the $r_{,v}=0$ horizons are time-like.
\item[(E)] If $\omega \lesssim -1.5$, we can find an initial condition that the $r_{,v}=0$ horizon shrinks to the singularity
and eventually a space-like $r_{,v}=0$ horizon and a singularity appears. We can see a Cauchy horizon.
\item[(F)] If $\omega \lesssim -1.5$, we also find an initial condition that the $r_{,v}=0$ horizon shrinks but not to a singularity.
Then, we see an oscillation of the $r_{,v}=0$ horizon.
\item[(G)] In $\omega \lesssim -1.5$ case, we can find an initial condition that the Cauchy horizon can be maintained infinitely.
\end{itemize}

\subsection{\label{sec:dis}Discussion}

Finally, we remark on some physical implications of our results.

First, in the $\omega \gtrsim -1.5$ limit, we could see a violation of the null energy condition for the $T_{uu}$ component.
This condition is a necessary condition to induce a bubble universe using a false vacuum bubble \cite{Yeom2}.
It is necessary to check further calculations to ascertain whether this is in fact useful;
but one lesson is that, in Brans-Dicke theory or similar type of theory,
we may apply the back-reaction of a gravitational collapse to induce the negative $T_{uu}$ part.

Second, in the $\omega \lesssim -1.5$ limit, we could see a violation of weak cosmic censorship.
Even though the Cauchy horizon extends infinitely, it is reasonable to speculate that a Schwarzschild black hole will be formed beyond the Cauchy horizon.
However, it is also true that we cannot determine what happens beyond the Cauchy horizon.
These situations were not observed using numerical simulations by previous authors \cite{Scheel:1994yr}.
Therefore, there should be further studies for this clear counterexample of weak cosmic censorship.

\section*{Acknowledgment}
The authors would like to thank Ewan Stewart, Alex Nielsen, Gungwon Kang, and Sungwook Hong for discussions and encouragement.
This work was supported by Korea Research Foundation grants (KRF-313-2007-C00164, KRF-341-2007-C00010) funded by the Korean government (MOEHRD) and BK21.

\section*{\label{sec:appa}Appendix A. Physical realizations for various $\omega$}

For observational tests, it is known that the value of $\omega$ should be greater than $4 \times 10^{4}$ \cite{Ber}. However, in various physical situations, small $\omega$ parameters can be allowed. Even though the small $\omega$ is not for our universe, if small $\omega$ is allowed in fundamental theory and may be realizable in somewhere of multiverse, and if such small value of $\omega$ has some implications, e.g., violation of unitarity or cosmic censorship, the study of responses of the Brans-Dicke field for small $\omega$ will have theoretical importance.

One example is the dilaton gravity which has the effective action as the following form \cite{Gasperini:2007zz}:
\begin{eqnarray}
\label{eq:dilaton} S = \frac{1}{2 \lambda_{s}^{d-1}}\int d^{d+1}x \sqrt{-g} e^{-\Phi} \left( R + (\nabla \Phi)^{2} \right),
\end{eqnarray}
where $d$ is the space dimensions, $\lambda_{s}$ is the length scale of string units, $R$ is the Ricci scalar, and $\Phi$ is the dilaton field. If we define $\phi$ as
\begin{eqnarray}
\label{eq:def} \frac{e^{-\Phi}}{\lambda_{s}^{d-1}} = \frac{\phi}{8 \pi G_{d+1}},
\end{eqnarray}
where $G_{d+1}$ is the $d+1$ dimensional gravitation constant, then we obtain the Brans-Dicke theory with $\omega = -1$ limit. If there are higher loop corrections from string theory, there will be other coupling terms of $\phi$ and hence the Brans-Dicke theory should be modified. Hence, the correspondence between the dilaton gravity and the Brans-Dicke theory is only for weak coupling limit ($e^{-\Phi} \gg 1$). However, it is reasonable to think that Brans-Dicke theory with $\omega=-1$ limit is a good toy model to study phenomena of dilaton gravity.

In the first model of Randall and Sundrum \cite{Randall:1999ee}, they introduced two branes to explain the hierarchy problem. Because of the warp factor between two branes, we obtain a positive tension brane and a negative tension brane in an anti de Sitter background. According to Garriga and Tanaka \cite{Garriga:1999yh}, each branes can be described by Brans-Dicke theory in the weak field limit with the $\omega$ parameter
\begin{eqnarray}
\label{eq:gt} \omega = \frac{3}{2} \left( e^{\pm s/l} - 1 \right),
\end{eqnarray}
where $s$ is the location of the negative tension brane along the fifth dimension, $l=\sqrt{-6/\Lambda}$ is the length scale of the anti de Sitter space, and the sign $\pm$ denotes the sign of the tension. To explain the hierarchy problem, we require $s/l \sim 35$. Then we obtain sufficiently large $\omega$ on the positive tension brane and $\omega \gtrsim -3/2$ on the negative tension brane. However, in principle, $s/l$ can be chosen arbitrarily and hence one may guess that various $\omega$ near $-3/2$ may be allowed by models from brane world.

For more discussions, see \cite{Fujii:2003pa}.

\section*{\label{sec:appb}Appendix B. Convergence and consistency tests}

In this appendix, we report on convergence and consistency tests for our simulations. We used $\omega=-1$, $A=0.75$, and $v_{\mathrm{f}}=20$ case.

\begin{figure}
\begin{center}
\includegraphics[scale=1]{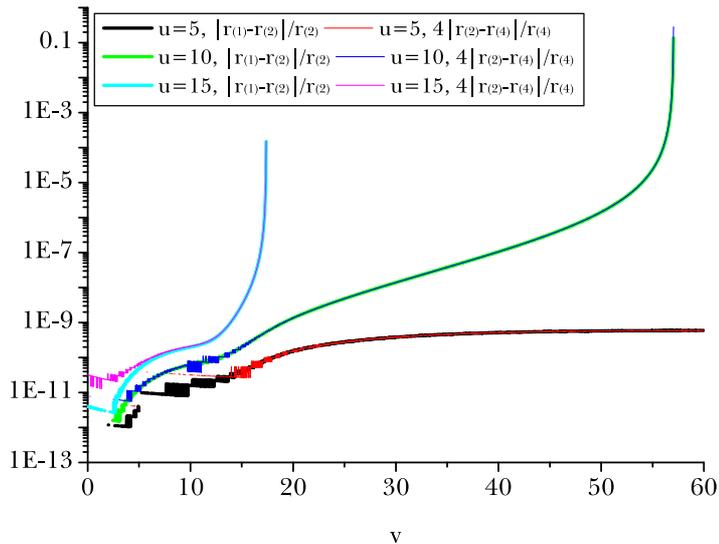}
\caption{\label{fig:convergence}Plots of errors with different step sizes.
Here, we plot $|r_{(1)}-r_{(2)}|/r_{(2)}$ and $4|r_{(2)}-r_{(4)}|/r_{(4)}$ along a few constant $u$ lines,
where $r_{(n)}$ is calculated in an $n\times n$ times finer simulation than $r_{(1)}$.
Since two curves are folded for each $u$ slice, we confirm that our simulation converges to second order.}
\end{center}
\end{figure}

For convergence, we compared finer simulations: $1\times1$, $2\times2$, and $4\times4$ times finer.
In Figure~\ref{fig:convergence}, we see that the difference between the $1\times1$ and $2\times2$ times finer cases is $4$ times the difference between the $2\times2$ and $4\times4$ times finer cases,
and thus our simulation converges to second order.
For $u=10$ and $u=15$, the curves appear to diverge, since there is a singularity. However, even before the singularity, the error is $\lesssim 0.1\%$.

\begin{figure}
\begin{center}
\includegraphics[scale=1]{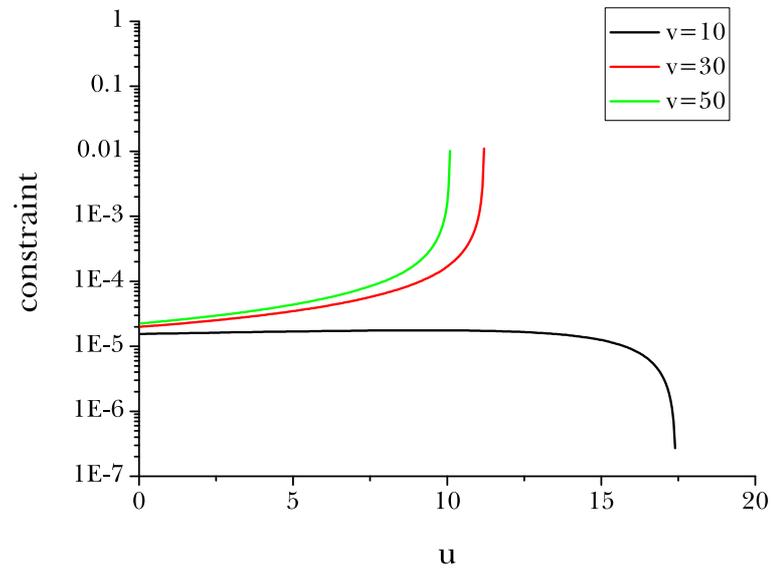}
\caption{\label{fig:constraint_u}Plot of Equation~(\ref{eq:constraint_u}) for $v=10, 30, 50$.}
\end{center}
\end{figure}
\begin{figure}
\begin{center}
\includegraphics[scale=1]{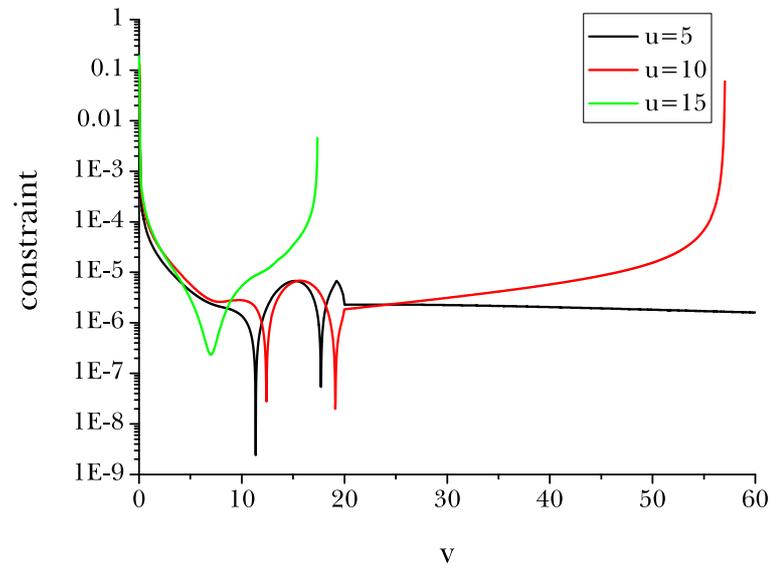}
\caption{\label{fig:constraint_v}Plot of Equation~(\ref{eq:constraint_v}) for $u=5, 10, 15$.}
\end{center}
\end{figure}

We also compared constraint equations (Equations~(\ref{eq:E1}) and (\ref{eq:E2})) using
\begin{eqnarray}
\label{eq:constraint_u}\frac{|f_{,u} - 2fh + \frac{r}{2 \phi} (w_{,u}-2hw) + \frac{r}{\phi}W^{2} + \frac{r \omega}{2 \phi^{2}} w^{2}|}
{|f_{,u}| + |2fh| + |\frac{r}{2 \phi}| (|w_{,u}|+|2hw|) + |\frac{r}{\phi}W^{2}| + |\frac{r \omega}{2 \phi^{2}} w^{2}|} &=& 0,\\
\label{eq:constraint_v}\frac{|g_{,v} - 2gd + \frac{r}{2 \phi} (z_{,v}-2dz) + \frac{r}{\phi}Z^{2} + \frac{r \omega}{2 \phi^{2}} z^{2}|}
{|g_{,v}| + |2gd| + |\frac{r}{2 \phi}| (|z_{,v}|+|2dz|) + |\frac{r}{\phi}Z^{2}| + |\frac{r \omega}{2 \phi^{2}} z^{2}|} &=& 0.
\end{eqnarray}
Figures~\ref{fig:constraint_u} and \ref{fig:constraint_v} show the constraint equations.
Around the singularity, the constraint equations increase, but still the value is less than $\sim1\%$.
For Figure~\ref{fig:constraint_v}, the constraint equation initially appears to be large,
because initially both the numerator and denominator, in Equation~(\ref{eq:constraint_v}), are too small (less than $\sim 10^{-10}$).
Therefore, these plots show that our simulations hold the constraint equations sufficiently.

\end{document}